\documentclass{mn2e}
\def\etal{{\it et al.\ }}
\def\eg{{\it e.g.,}}
\def\ie{{\it i.e.,}}

\begin{document}

\title[Interactions as star formation triggers]{Are interactions the primary
triggers of star formation in dwarf galaxies?}

\author[N. Brosch et al.]{Noah Brosch$^1${\thanks{noah@wise.tau.ac.il}},
Elchanan Almoznino$^1$, Ana B. Heller$^1$\\
$^1$The
Wise Observatory and the School of Physics and
Astronomy, the Raymond and  Beverly Sackler Faculty of Exact
Sciences, \\ Tel Aviv University, Tel Aviv 69978, Israel}

%\date{Revised Draft: 27 November 2003}

\maketitle

\begin{abstract}

We investigate the assumption that the trigger of star formation
in dwarf galaxies are interactions with other galaxies, in the
context of a search for a ``primary'' trigger of a first
generation of stars. This is cosmologically relevant because the
galaxy formation process consists not only of the accumulation of
gas in a gravitational potential well, but also of the triggering
of star formation in this gas mass and also because some high-z
potentially primeval galaxy blocks look like nearby star-forming
dwarf galaxies. We review theoretical ideas proposed to account
for the tidal interaction triggering mechanism and present a
series of observational tests of this assumption using published
data. We show also results of a search in the vicinity of a
composite sample of 96 dwarf late-type galaxies for interaction
candidates showing star formation. The small number of possible
perturbing galaxies identified in the neighborhood of our sample
galaxies, along with similar findings from other studies, supports
the view that tidal interactions may not be relevant as primary
triggers of star formation. We conclude that interactions between
galaxies may explain some forms of star formation triggering,
perhaps in central regions of large galaxies, but they do not seem
to be significant for dwarf galaxies and, by inference, for
first-time galaxies forming at high redshifts. Intuitive
reasoning, based on an analogy with stellar dynamics, shows that
conditions for primary star formation triggering may occur in gas
masses oscillating in a dark matter gravitational potential. We
propose this mechanism as a plausible primary trigger scenario,
worth investigating theoretically.

\end{abstract}

\begin{keywords}
galaxies-dwarf, star formation, interactions, trigger
tides
\end{keywords}

\section{Introduction}

Many galaxies show detectable star formation (SF) events. These
can have different levels of intensities, from very strong SF
events called ``starbursts'' to minor, almost indistinguishable SF
processes, and can occur either in the centers of galaxies or in
their outer regions. At the high end of the intensity scale one
finds ultraluminous IR-emitting galaxies (ULIGs: Soifer \etal
1987), probably giant and strongly interacting galactic systems.
Most of the radiation emitted by the newly formed stars is
re-emitted as IR radiation, because their interstellar matter
(ISM) is loaded with high quantities of dust. The SF taking place
in ULIGs is located primarily in their central regions. Another
type of star-forming galaxies look almost like stars but show
copious SF; these are the compact SF objects.

At the faint end of the SF scale one finds low surface brightness
(LSB) dwarf irregular galaxies (DIGs). In these systems the SF is
not nuclear or systemic, but one finds small SF regions scattered
over the galaxy's surface (Heller \etal 1999). In between these
extremes, one encounters various levels of SF in most late-types
of galaxies.

The nature of the SF triggering in galaxies has been studied by
many and a plethora of mechanisms were proposed. In many types of
galaxies, more than one mechanism probably operate. In large,
well-organized disks there are global SF triggers, such as density
waves and shear forces produced by differential rotation. These do
not operate in DIGs, which rotate mostly as solid bodies or are
supported by turbulent motions, and are generally devoid of spiral
structure. The reduction in the number of possible SF triggers
implies that DIGs are good subjects in which to study the SF
phenomenon in a simple environment.

The more intense the SF, and the longer an SF episode lasts, the
more metals are produced in a galaxy. One, therefore, hopes to
understand the process of galaxy formation at high redshifts by
studying it in nearby, metal-poor dwarf galaxies. The three
extragalactic objects with the lowest known metallicity are all
dwarf galaxies (I Zw 18, SBS 0335-052, and SBS 0822+3542) and were
proposed at times as ``first-time galaxies''. However, the
analysis of the underlying stellar population of I Zw 18 (Aloisi
\etal 1999) indicates that this is not a first-time galaxy,
despite its paucity of metals. I Zw 18 has probably been forming
stars at a slow and steady rate over at least one Gyr.

We investigated SF and stellar populations in two samples of DIGs
in the Virgo cluster (Almoznino \& Brosch 1998a, 1998b; Heller
\etal 1999). We showed that the high surface brightness DIGs of
the ``blue compact dwarf'' (BCD) flavor can be understood as
originating from a number of starburst events. Each such starburst
episode is relatively short, lasts only a few Myrs, and the
episodes are separated by a few hundred Myrs up to a few Gyrs. We
are currently investigating  in a similar manner samples of SF
 galaxies in voids.

We studied the morphology of star formation in DIGs, hoping that
the distribution of HII regions identified as loci of recent SF
may pinpoint the proper SF triggering mechanism. The connection
between the asymmetry of a galaxy and its star formation
properties has already been established by \eg \, Conselice,
Bershady \& Jangren (2000), but not for a sample of DIGs. We
showed that the SF in DIGs is asymmetric and takes place
preferentially far from the galaxy's center, more to one side of a
galaxy (Brosch \etal 1998). We also showed (Heller \etal 2000)
that this pattern of SF is present not only in Virgo DIGs, but
also in other samples of BCD and LSB dwarfs, and that it is
reflected not only in the properties of the number counts of HII
regions, but also in the distribution of the H$\alpha$ flux over
the galaxy. Therefore, the asymmetry and off-center SF results are
not biased by the small-number statistics of HII regions (Poisson
asymmetry). The validation of these results for objects outside
the Virgo cluster demonstrates that the cluster environment is not
responsible for the asymmetric SF property of DIGs, in line with
our preliminary conclusion about the lack of dependence of the
H$\alpha$ equivalent width in Virgo DIGs on the galaxies'
(projected) Virgocentric distance (Heller \etal 1998). Similar
results, on the location of sites of recent SF in ten nearby DIGs,
have been reported by Schulte-Ladbeck \& Hopp (1998).

Here we concentrate on the possibility that star formation in
dwarf irregular or compact (BCG) galaxies is triggered by tidal
interactions with other nearby galaxies. We do this mainly by
searching for star-forming companion galaxies in a large composite
sample of DIGs, all imaged in H$\alpha$. The lack of such
companions for most of the sample galaxies, along with other
supporting evidence, indicates that this mechanism of
star-formation triggering by external interactions can also
probably be discounted as a primary SF trigger. A primary trigger,
in our case, is understood to be one that does not require any of
the previously-mentioned factors in order to become active. Such a
trigger could operate in isolated blue compact galaxies, where
even a tidal interaction with another dwarf galaxy is probably not
sufficient to trigger a SF event (cf. Kunth \& \"{O}stlin 2000).
Note also that our negative results could, in principle, also
indicate that the SF in our sample galaxies is entirely secondary,
\ie \, triggered by internal sources.

We emphasize that the search here is for plausible primary SF
triggers that could initiate a first generation of stars in a
galaxy. Some of the DIGs we study are good candidates for
``first-time'' galaxies. Once some massive stars have formed in a
first SF burst, their subsequent evolution produces strong stellar
winds and supernova shocks. These may trigger secondary SF, as
assumed in the stochastic self-propagating star formation
mechanism (Gerola \& Seiden 1978). Stewart (1998) showed that this
secondary SF mechanism operates in the dwarf galaxies Ho II, IC
2574, and Sextans A. However, searches for primary SF triggers
have, so far, been unsuccessful ({\it e.g.,} van Dyk \etal 1998,
for Sextans A).

The plan of this paper is to present in section 2 published claims
that interactions trigger SF, then to explain in section 3 one of
the proposed mechanisms by which SF could be triggered by
long-range interactions, and to examine in section 4 the new
observational evidence, which is mainly against this possibility.
The discussion in section 5 will focus on this negative evidence
and a scenario will be proposed by which primary SF could be
internally triggered.

\section{Claims that interactions trigger star formation}

The field of galaxy formation and evolution has seen many claims
that tidal interactions trigger SF, but very few clear-cut tests
of these claims, or detailed theoretical models of such
interactions, have ever been presented. We show below that tidal
interactions could be responsible for triggering nuclear SF, but
they cannot be invoked blindly as a general SF trigger (\eg \,
Hutchings \etal 1999).

Melnick (1987) proposed that collisions between intergalactic gas
clouds and dwarf galaxies trigger star formation in the latter.
The problem with this assumption is the apparent lack of
significant numbers of intergalactic gas clouds, at least in the
form of HI. At about the same time, Noguchi (1987) suggested that
interactions between galaxies can drive ISM to the centers of
galaxies through the formation of a bar, causing nuclear
starbursts to take place. Brinks (1990) suggested that collisions
between DIGs may trigger SF. Olson \& Kwan (1990a, b) suggested
that interactions between galaxies drive ISM cloud-cloud
collisions within a galaxy, which may trigger bursts of SF.

Taylor \etal (1995) detected more HI companions around HII
galaxies (dwarf irregular galaxies with strong H$\alpha$ emission)
than around quiescent LSB dwarfs; these could be the clouds
suggested by Melnick (1987), but they also remarked that a sizable
fraction of the HII galaxies do not appear to have companions at
all. Taylor (1997) argued that the excess of companions around HII
galaxies may indicate that these are LSBs where SF has been
activated by interactions. Telles \& Terlevich (1995) showed that
there are more bright (non dwarf) galaxies within 1 Mpc$^3$ of HII
galaxies than expected from a random distribution. Also, Pisano \&
Wilcots (1999) searched for HI companions around six extremely
isolated galaxies from the Nearby Galaxies Catalog (Tully 1988).
They found that two of the sample galaxies are really the primary
members of triple systems where the HI companions are within 90
kpc of their primary galaxy, but that the other four galaxies
appeared isolated in that they did not show HI neighbors.

Moss \etal (1998) searched for H$\alpha$ emission in galaxies near
the Abell 1367 cluster. They found that objects with compact
H$\alpha$ emission tend to show a disturbed morphology. From this,
they concluded that compact line emission, \ie \, nuclear star
formation, results from tidally-induced SF.

Telles \& Melnick (1999) searched for faint companions near HII
galaxies using APM plate scans. They concluded that although HII
galaxies appear significantly clustered, because of a positive
cross-correlation with the APM galaxies, their SF could not be
triggered by tidal interactions. The finding of Loveday \etal
(1999) that the majority of local dwarf galaxies undergo star
formation at present, and that they are preferentially found in
low-density environments, also argues against galaxy interactions
playing a role in promoting SF. On the other hand, the disturbed
appearance of many high-redshift galaxies ({\it e.g.,} objects in
the MDF studied by Abraham \etal 1996; galaxies in the Hubble Deep
Field considered by Conselice \& Bershady 1998) has been explained
as signs of mergers and interactions.

Iglesias-P\'{a}ramo \& Vilchez (1999) checked for differences in
the SF rate between disk galaxies in compact groups and field
galaxies. They found that the only characteristic that seems to be
different is the width of the SFR distribution, which is broader
for the compact group sample. This seems to indicate that the
dynamic state of a group does not influence the SFR of a galaxy.

Carter {\it et al.} (2001) studied the star formation in a large
sample of nearby (z$\approx$0.05) galaxies. They concluded that
the presence of star formation in a galaxy is strongly correlated
with the neighborhood galaxy density, defined as the density over
a few Mpc. They also found that the stellar birthrate parameter
\begin{equation}\label{StellerBirthrateParam}
    b=\frac{SFR}{<SFR>}
\end{equation}
does not correlate with the local galaxy density, from which they
inferred that the star formation triggering takes place on a
smaller spatial scale than the one defining the Mpc-scale
neighborhood. Similar conclusions were reached by Noeske \etal
(2001) from a study of 98 star-forming dwarf galaxies.
Specifically, they searched for companions around these galaxies
using NED\footnote{The NASA/IPAC ExtragaSFlactic Database (NED) is
operated by the Jet Propulsion Laboratory, California Institute of
Technology, under contract with the National Aeronautics and Space
Administration.}and failed to detect significant differences with
distance from the companion in either of two signs of recent SF:
the H$\beta$ equivalent width, or the blueness of the B-V color of
the dwarf galaxy.

Schmitt (2001) used galaxies in the Palomar survey with
$B_T\leq$12.5 mag in the northern hemisphere to test for the
presence of activity (Seyfert, HII, LINER, etc.) as a function of
the local galaxy density or the presence of companions in the NED
and the Digitized Sky Survey. He found, contrary to previous
studies, that Seyfert and HII galaxies tend to have fewer
companions than other types of galaxies.

An imaging Fabry-Perot study of a small sample of Blue Compact
Galaxies (BCG) by \"{O}stlin \etal (2001) indicated that star
formation there is likely to be triggered by mergers with gas-rich
dwarf galaxies or by massive gas clouds. Severgnini \& Saracco
(2001) argued that in the Hickson Compact Groups (HCGs) the
correlation between the H$\alpha$ luminosity of galaxies on the
one hand, and the velocity dispersion of a group and its
compactness on the other, implies that SF takes place where there
is a higher probability of galaxy interaction (lower velocity
dispersion for the group).

Pustilnik \etal (2001a) studied the environment of a sample of 86
compact emission-line galaxies and found that about one-third are
companions of significantly brighter galaxies. Another $\sim$1/3
have companions of similar brightness or fainter, and about 15\%
show indications of previous mergers. From this, they concluded
that SF bursts are triggered by interactions with other galaxies.

In a more detailed work based on the same galaxy sample, Pustilnik
\etal (2001b) separated their sample into one sub-sample that
belongs to the "Local Supercluster" and one that represents the
"general field" and found small but insignificant differences
between galaxies in the two environments in the aspect of the
presence of bright disturbing galaxies. Among the galaxies lacking
bright disturbers, ``isolated'', as per Pustilnik \etal (2001b),
they claim that ~43\% have dwarf disturbers and that ~26\% show a
merger morphology. These claims will be examined again below. Note
that Pustilnik \etal (in their various papers) use an SF
triggering mechanism that could activate the process through
``soft'' long-range interactions proposed by Icke (1985); this
mechanism and its relevance to dwarf galaxies will be discussed
below.

Barton Gillespie \etal (2003) studied the SF properties of 190
galaxies in pairs or in compact groups. Their sample represents
nearby bright objects (m$_{Zw}\leq$15.5; $v\leq$2,300 km
sec$^{-1}$) and the galaxies are relatively close to each other
(projected distance $\Delta D\leq50 h^{-1}$ kpc; $\Delta v \leq$
1,000 km sec$^{-1}$). They found a clear indication for
interaction-triggered SF, with the strongest SF bursts occuring in
galaxies in the tightest orbits. Barton Gillespie \etal found also
that low-mass galaxies experience stronger SF bursts than massive
galaxies.

Finally, Hoffman \etal (2003) studied the HI content of a small
sample of Virgo cluster BCDs with Arecibo and VLA D-array mapping.
These galaxies show, by definition, strong on-going SF bursts.
Their HI maps showed that in about 1/3 of the galaxies no signs
were found for companion galaxies or for gas clouds that may have
triggered the observed SF event.

In principle, another type of interaction could also act to
trigger SF. This is a tidal triggering by the cluster
gravitational potential in which one of the scenarios (Henriksen
\& Byrd 1996) produces a significant lateral compression of a
galaxy disk. This causes enhanced cloud-cloud collisions in the
disk, followed by SF, but can happen only for rather close
passages near the cluster center (within 250 kpc) and immediately
following the passage (within 1/4 of a disk rotation period). This
mechanism should, therefore, only enhance SF for galaxy disks near
a cluster center. A similar conclusion was reached by Fujita
(1998), while Gnedin (2003) showed that dSph and LSB galaxies are
unlikely to survive cluster tides. Observationally, the survey by
Moss \& Whittle (2000) showed that, at least for spiral galaxies,
the tidal interactions cause mainly circumnuclear SF, not the
diffuse type of starbursts encountered in some dwarf galaxies. For
these reasons we believe that the trigger from cluster tides can
be discounted for the case of dwarf galaxies.

The conclusion from the evidence summarized above is that the
situation regarding SF triggering by interactions is, at best,
confusing. There seems to be no clear-cut case of tidal triggering
of star formation as a general phenomenon. In many cases, the
triggering body that should have had a close encounter with the
star-forming galaxy seems to be missing.

\section{The Icke star formation trigger mechanism}

Could it be possible that SF may be triggered by distant
encounters with other galaxies, so that we may not recognize such
an event because the galaxies involved are too far apart at
present? Following the formalism of Icke (1985), SF could be
induced in a DIG by an interaction with another small galaxy, as
suggested by Brinks (1990). Icke considered distant encounters
between disk galaxies as possible triggers of SF and as a possible
explanation for the existence of S0 spirals. His proposal is based
on relatively simple numerical hydrodynamic calculations of a gas
disk perturbed by an intruder galaxy passing by on a hyperbolic
planar orbit. Icke showed that, in some instances, shocks may be
generated in the disk even in situations when tidal bridges or
tails will not form.

Briefly, Icke's (1985) argument requires a close passage in a
prograde sense between the ``victim" galaxy and the ``intruder"
galaxy in order to produce shocks in a gas disk. This would speed
up the gas by $s_0$ (the sound speed in the ISM gas) in about half a
rotation period of the victim. A supersonic shock is then assumed
to trigger SF. The maximal peri-galactic distance between the
intruder and the victim, in order that shocks will take place, must
be
\begin{equation}
p_0\approx(\frac{8 \, \pi \, \mu}{s_0})^{1/3}
\end{equation}
where $\mu$ is the ratio
of the intruder mass to the mass of the victim.

Icke's (1985) argument was designed for large spiral disks as
victims and large galaxies as intruders, and was aimed to explain
S0 galaxies. Its importance is in providing an apparent way to
form stars by galaxy-galaxy interactions that do not leave visible
``traces'' in the form of tidal tails or direct collisions.
Assuming that this argument holds also for dwarf galaxies, one has
only to change the mass ratio and evaluate $p_0$ in units of
$r_0$, that is, in units of the disk scale length of the victim
DIG. It is possible to use luminosity ratios instead of mass
ratios for the two interacting galaxies by assuming some
mass-to-light values. If the mass-to-light value is 8 (a somewhat
random choice, typical for DIGs and close to the ``maximum disk"
value: Gerritsen \& de Blok 1999), we find that for a difference
of 3 mag between the DIG victim and the large galaxy intruder the
peri-galactic passage should be closer than $\sim56 r_0$ to
trigger an SF shock. A magnitude difference of 6 mag implies $p_0
\leq 140  r_0$ for shocks to be produced.

The implication is that coplanar prograde peri-galactic passages
closer than 100-200 kpc (for a typical scale length of 1 kpc for a
DIG) should cause ``Icke shocks" that could be responsible for
enhanced SF in dwarf galaxies, whereas more distant or out-of-plane passages could
not cause such shocks. Note that the delay between the
peri-galactic passage of a victim to the onset of shock conditions
(one half of the victim's rotation as claimed by Icke) does not
significantly alter the distance between the galaxies. Larger
values of $p_0$ are feasible for DIGs with $r_0\gg$1 kpc, or with
M/L$\gg$8, or with both. It is likely that even if one of these
assumptions is fulfilled the victim will not be a typical DIG. For
instance, the faint, compact narrow emission line galaxies studied
by Guzm\'{a}n \etal (1998) at z$\sim$0.22-0.66 are quite luminous
at M$_B \approx$--21, have scale lengths of 1-5 kpc, yet have the
masses of dwarf galaxies ($\leq10^{10}$ M$_{\odot}$).

\section{Observational evidence}

In this section we first review published observations that can be
used to test the idea of interaction-triggered star formation,
then examine samples of dwarf galaxies studied previously to
understand their star-forming properties but concentrating now on
finding possible star-forming physical companions.

\subsection{Searches for companions in published samples of galaxies}

It is possible to test observationally the relation from Icke
(1985) adapted to DIGs by considering the sample of dwarf galaxy
companions of large galaxies from Zaritsky \etal (1997). The
sample consists of 115 satellites of 69 ``primary'' galaxies, which
are closer than 500 kpc (projected) and have a velocity difference
from that of the parent galaxy smaller than 500 km sec$^{-1}$. The
satellites are at least 2.2 mag fainter than their parent galaxy.
The parent galaxies are isolated, in the sense that all companions
within 1,000 km sec$^{-1}$ and 500 kpc projected distance are
fainter by more than 2.2 mag, and those within 1,000 km sec$^{-1}$
and between 0.5 and 1.0 Mpc are at least 0.7 mag fainter than the
primary galaxy. The sample is useful, therefore, to study the
frequency of SF in the companions as a general property and as a
function of their distance from the "intruder", that is, the
primary large galaxy.

Zaritsky \etal (1997) list, for each of the companions, the
presence or absence of emission lines in their spectra. Although
not specifically mentioned in the paper, it seems that the lines
used for establishing the presence or absence of emission were the
[OII] doublet, H$\beta$, and the [OIII] lines. We take the
presence of emission lines to imply the existence of young,
massive stars that signal the presence of a recent SF event in the
dwarf companion. The inference, if Icke's (1985) proposed SF
trigger mechanism can be applied here, is that the closer a dwarf
is seen to its parent galaxy, the ``intruder" in Icke's
terminology, the more likely would it be to exhibit emission lines
in its spectrum. In particular, by binning the companions into
``close" and ``distant" groups, we expect to observe a marked
difference between them in terms of their SF properties.

We counted in the list of Zaritsky \etal (1997) the number of
companions with emission lines within 200 kpc projected
separation, and the number within 200 to 400 kpc. We found that
exactly 85\% of the companions in both cases exhibited emission
lines (47 out of 55 for the companions closer than 200 kpc, and 33
out of 39 for those more distant). It seems that the presence of
emission lines is a general characteristic of the companions of
large galaxies. We conclude that the assumption of tidally-induced
SF, manifested as a result that the closer sub-sample would form
stars more readily thus would show a higher fraction of
star-forming dwarfs, was not satisfied. This supports the
assertion that the Icke (1985) mechanism is probably not relevant
as a trigger of SF in dwarf companions of large galaxies where
tidal SF triggering could be expected.

On the other hand, Pustilnik \etal (1997) also claimed that many
blue compact galaxies (BCGs) have bright, nearby companions. They
used objects from the Second Byurakan Survey (SBS) and searched
their neighborhoods for brighter objects with which the BCGs could
have interacted. The Pustilnik \etal sample consists of 62 BCGs
with $v_{\odot}\leq6,000$ km sec$^{-1}$, from which they selected
a sub-sample of 26 BCGs closer than 2,500 km sec$^{-1}$. The
possible companions were selected to be brighter by at least 1.4
mag than the target BCG and their radial velocities were checked
spectroscopically. The search strategy was, therefore, opposite
that adopted by Zaritsky \etal (1997). Pustilnik \etal searched
for companions of the BCGs that are brighter than the BCGs.
Objects within 300 km sec$^{-1}$ were adopted as companion
galaxies, \ie \, possible ``intruders", following Icke's
classification.

The reported success rate in finding intruders was 15/26 (58\%)
for the nearby sub-sample, and 28/62 (45\%) for the entire BCG
sample. If one assumes Poisson statistics for the likelihood of
detection, these percentages are not different and indicate that
about half of the galaxies have companions. The mean projected
separation of a ``victim" BCG from its brighter, presumably more
massive ``intruder" was $\sim$250 kpc, and the mode of the
distribution of the magnitude differences between the two objects
was $\sim$3 mag. From this, Pustilnik \etal (1997) concluded that
SF activity in BCGs is most likely triggered by long-range tidal
interactions with large neighbors, following Icke's (1985)
proposal. However, it appeared that closer passages do not,
necessarily, trigger more intense or more frequent SF (cf. also
Carter \etal 2001).

The findings of Pustilnik \etal (1997) contradict, apparently, the
evidence from the data of Zaritsky \etal (1997). Note first that
the terms ``BCG'' and ``BCD'' are not interchangeable. The 1.4 mag
difference between the victim and the intruder does not ensure
that the BCG victim would always be fainter than M$_B\approx$--19
so as to qualify as a dwarf galaxy. It is possible that many BCG
sample galaxies of Pustilnik \etal are not dwarf galaxies. It is
also likely that the SF they detected in the sample objects is in
many cases nuclear, and this could be related to tidal
interactions. A hint that this could be the case is that the
galaxies are SBS objects, \ie \, were detected as compact
emission-line objects in an objective-prism survey, while some of
them appear extended in wide-band images.

\subsection{Star formation triggering in small galaxies}

We assume that if the SF is triggered by an interaction with a
neighboring small galaxy, the trigger should operate not only on
the target galaxy but also simultaneously on the disturbing
object. Therefore, we assume in what follows that when H$\alpha$
emission is triggered in the DIG that is the target of the SF
study, similar H$\alpha$ emission appears in (at least) one of the
galaxies in the immediate neighborhood of the target DIG. This
assumption may not be fulfilled if the disturbing galaxy has no
interstellar gas, thus cannot form new, massive stars, or if it
does not have sufficient ISM to be ionized and be detected as
H$\alpha$ emission, or if the SF timescales in the two objects are
widely different.

We studied a composite sample of 96 DIGs, most of which were used
initially to explore the issue of asymmetric and non-central SF in
such objects (Heller \etal 2000). The galaxies originate from the
Almoznino \& Brosch (1998) studies of Virgo BCDs extended to
include all the Virgo galaxies classified as BCD, or any other
morphological class and BCD, (N=40; Brosch \etal in preparation),
and from the Heller \etal (1999a) study of a complete Virgo LSB
DIG sample (N=18). As the original studies required that the
galaxies be members of the Virgo cluster (VC), we imposed also the
following selection criteria: (a) a positive detection in HI, as
listed in Hoffman \etal (1987, 1989), and (b) a radial velocity
from optical or HI observations smaller than 2600 km s$^{-1}$ to
ensure membership in the VC. Most galaxies in Heller \etal are
classified in NED as ImIV or ImV.

We added galaxies from a private collection of Liese van Zee (LvZ;
N=18), and from a similar collection by John Salzer (JS; N=20).
The latter two lists were already used in our study of SF
asymmetry (Heller \etal 2000). Because of this non-uniform
selection, the composite sample is not representative or complete
in any way; it is used it here only as an indicative survey of SF
properties among companions of dwarf galaxies. For instance, while
most LvZ galaxies are classified in NED as Magellanic irregulars
(Im), those of JS are mostly BCDs. The fraction of objects
carrying the classification BCD from the entire sample is 64\%
(61/96).

The galaxies in the composite sample are uniformly classified as
dwarf galaxies, of subclasses BCD, LSB, or Im of various subtypes.
The objects are all relatively nearby, at distances of order that
of the Virgo cluster or nearer, with only two objects (UGC 2535
and UM 408) at redshifts higher than 2600 km sec$^{-1}$. They all
have net H$\alpha$ images as well as continuum light images
obtained in an off-H$\alpha$ band or in the R band, but in rare
cases only in the B band. The net-H$\alpha$ images were obtained
by subtracting the continuum contribution from the rest-frame
H$\alpha$ image. The red continuum image was scaled to the
H$\alpha$ image by a factor chosen to cause stars to fully
subtract from the net image. This procedure is justified, because
the H$\alpha$ line filters used for rest-frame galaxy photometry
sample line-free segments of the spectra in most stars.

The net H$\alpha$ line images were inspected visually for the
presence of diffuse objects in the field, not obviously related to
the target galaxy (outer HII regions in the galaxy, for example).
Note that the images from the LvZ list have had some of the
brighter stars and galaxies blocked off in the line image; this
could have masked off some H$\alpha$ contribution eliminating some
possible companions. As the H$\alpha$ filters used in this survey
are all rather narrow, any such diffuse image detected in the
net-H$\alpha$ image represents a galactic-shaped object at
$\sim$the same redshift as the target galaxy. Specifically, the
typical filter FWHM of 50\AA\, implies a maximal redshift
difference of $\sim$2,000 km sec$^{-1}$ between the target galaxy
and a potential companion if the two galaxies are located at the
two ends of the transmission band. In reality, the redshift
difference must be much smaller in order for a potential companion
to be revealed in the net line image, because the transmission
profile of the filter would attenuate significantly the intensity
of an H$\alpha$ line originating from an object red- or
blue-shifted close to the edge of the filter band and, in most
cases, the central wavelength of a filter would be tuned to the
target galaxy redshift. The visual inspection is limited to the
size of the net H$\alpha$ image, and the projected size of the
search area changes with the distance to the target galaxy.

Note that, unlike other studies of this topic, we do not claim
that a disturbed morphological appearance is a clear indicator of
an interaction. This criterion might be valid for large,
well-behaved galaxies that have regular patterns, but it is not
acceptable for dwarf irregular galaxies. The presence of
companions was verified by blinking the line and continuum images;
real companion galaxies must have counterparts in both images
(line and continuum), ruling out the detection of galaxies with
very dim continua and strong emission lines. In principle, the
non-H$\alpha$-emitting galaxies could also be ``intruder''
galaxies and may have triggered SF if they are at the right
redshift, but this could not be detected with the present
technique.

We also conducted a NED search for possible nearby companions,
limiting this to a projected distance of five arcmin and a
$\Delta$v of 1000 km sec$^{-1}$. For a typical distance of 20 Mpc
to the sample galaxies, this angular distance corresponds to a
projected distance of 300 kpc. The companions found this way are
listed in the remarks column of Table 1 as 'pair with' (p.w.),
followed by the name of the companion and its approximate
projected distance in arcmin.

\begin{table*}
\begin{minipage}{150mm}
\caption{Possible HII companions of dwarf star-forming
galaxies: Virgo BCDs}
%\label{symbols}
\begin{tabular}{@{}ccccccl}
\hline
Name & Type & v$_{\odot}$ (km sec$^{-1}$) & a (') & A (a) & N & Remarks \\ \hline
%NAN extended  \\ %\nl using the grey scale plots from thesis

VCC 0010 & BCD: & 1971 & 0.6 & 2 & 0  \\ %\nl

VCC 0022 &  BCD? & 1691 & 0.1 & 60 & 0  \\ %\nl

VCC 0024 & BCD & 1289 & 0.5 & 24 & 0  \\ %\nl

VCC 0130 & BCD? & 2189 & 0.3 & 30 & 0  \\ %\nl

VCC 0135 & S pec/BCD & 2378 & 1.0 & 10 & 0  \\ %\nl

VCC 0144  & BCD & 2021 & 0.4 &  3 & 0  \\ %\nl

VCC 0172 &  BCD: & 2175 & 0.7 & 20 & 0 &    \\ %\nl

VCC 0207 & BCD? & 2564 & 0.25 & 45 & 0  \\ %\nl

VCC 0213  & dS?/BCD  & --154 &  0.6 & 15 & 0 &  \\ %\nl

VCC 0223 & BCD? & 2070 & 0.2 & 50 & 0  \\ %\nl

VCC 0281 & dS0 or BCD &  257 & 0.4 & 53 & 0 & IC 3120  \\ %\nl

VCC 0309 & Im/BCD & 1566 & 0.6 & 19 & 0  \\ %\nl

VCC 0324 & BCD & 1526 & 0.7 & 14 & 0  \\ %\nl

VCC 0334  & BCD & --254 & 0.3 & 70 & 0  \\ %\nl

*VCC 0340  & BCD or merger & 1512 & 0.5 & 18 & 0 & in WBL 392
 \\ %\nl

VCC 0410 & BCD & 283 & 0.2 & 54 & 0 & RMB 12669  \\ %\nl

VCC 0428 & BCD & 794 & 0.4 & 77 & 0 & p.w. VCC 0413@4'.2 \& $\Delta$v=520 km sec$^{-1}$ \\ %\nl

VCC 0446 & Im/BCD: & 825 &  0.8 & 11 & 0 &   \\ %\nl

VCC 0459 & BCD & 2107 & 0.4 & 24 & 0 &   \\ %\nl %H$\alpha$ image in NED

VCC 0468 & BCD? & 1979 & 0.3 & 27 & 0  \\ %\nl

VCC 0513 & BCD? & 1832 & 0.4 & 16 & 0   \\ %\nl

VCC 0562 & BCD & 44 & 0.3 & 3 & 0 & KUG1220+124  \\ %\nl

VCC 0641 & BCD? & 906 & 0.4 & 15 & 0  \\ %\nl

VCC 0655 & S pec, N:/BCD & 1146 & 1.7  & 11 & 0 & NGC 4344  \\ %\nl

VCC 0737 & Sd/BCD? & 1725 & 1.0 & 9 & 0  \\ %\nl

VCC 0741 & BCD? & 1861 & 0.4 & 17 & 0  \\ %\nl

VCC 0772 & BCD? & 1226 & 0.3 & 29 & 0  \\ %\nl

VCC 0841 & BCD & 501 &  0.4 & 24 & 0 & RMB 46  \\ %\nl

VCC 0848 & ImIII pec/BCD & 1537 & 1.0 & 15 & 0  \\ %\nl

VCC 0890 & BCD? & 1483 & 0.14  & 28 & 0  \\ %\nl

VCC 0985 & BCD? & 1638 &  0.3 & 25 & 0 &  \\ %\nl

VCC 1141 & BCD? & 1040 & 0.32 & 55 & 0 &  p.w. VCC 1164@3'.4 \& $\Delta$v=0 km sec$^{-1}$  \\ %\nl

VCC 1179 & ImIII/BCD & 764 & 1.0 & 15 & 0 & IC 3412  \\ %\nl

VCC 1313 & BCD & 1254 & 0.2 & 18 & 0 & RMB 132  \\ %\nl

VCC 1356 & SmIII/BCD & 1251 & 0.6 & 14 & 0 & IC 3446  \\ %\nl

VCC 1374 & ImIII/BCD & 2559 & 1.0 & 12 & 0 & IC 3453  \\ %\nl

VCC 1437 & BCD & 1160 & 0.3 & 21 & 0  \\ %\nl

VCC 1459 & BCD: & 1774 & 0.4 & 4 & 0 &  p.w. IC 3474@4'.2  \& $\Delta$v=47 km sec$^{-1}$  \\ %\nl

VCC 1572 & BCD & 1848 & 0.5 & 34 & 0  \\ %\nl

VCC 1725 & SmIII/BCD & 1067 & 1.1 & 10 & 0  \\ %\nl

VCC 1744 & BCD & 1150 & 0.36 & 43 & 0 & p.w. IC 3602@5' \& $\Delta$v=129 km sec$^{-1}$  \\ %\nl

VCC 1750 & BCD? & --117 & 0.2 & 13 & 0 &   \\ %\nl

VCC 1791 & SBmIII/BCD & 2088 &  1.4 & 9 & 0 & IC 3617  \\ %\nl

VCC 1804 & ImIII/BCD & 1898 & 0.9 & 35 & 0 &  \\ %\nl

VCC 1955 & S pec/BCD & 2012 & 1.2 & 10 & 0 & NGC 4641  \\ %\nl

VCC 2007 & ImIII/BCD: & 1857 & 0.4 & 28 & 0 & IC 3716  \\ %\nl

VCC 2033 & BCD & 1486 & 0.4 & 51 & 0 &  \\ %\nl

VCC 2037 & ImIII/BCD & 1142 & 1.3 & 20 & 0 & p.w. VCC 2034@3' \&
$\Delta$v=459 km sec$^{-1}$
 \\ %\nl
\end{tabular}
\end{minipage}
\end{table*}

%\hline
%AnaHeller
\begin{table*}
\begin{minipage}{150mm}
\caption{Possible HII companions of dwarf star-forming
galaxies: Virgo LSBs}
%\label{symbols}
\begin{tabular}{@{}ccccccl}
\hline
Name & Type & v$_{\odot}$ (km sec$^{-1}$) & a (') & A (a) & N & Remarks \\ \hline

VCC 0017 & ImIV & 826 & 1.3 & 3 & 0 & U7150  \\ %\nl

VCC 0168 & dE2 or ImIV & 682 & 0.4 & 5 & 0  \\ %\nl

VCC 0169 & ImV & 2222 & 0.8 & 3 & 0  \\ %\nl

VCC 0217 & ImIV-V: & 1184 & 1.7 & 2 & 0 & U7307  \\ %\nl

VCC 0260 & ImIV & 1775 & 0.6 & 5 & 0  \\ %\nl

VCC 0328 & ImIV & 2179 &  0.9 & 15 & 0 & Boe 113  \\ %\nl

*VCC 0329 & ImV? & 1622 & 0.3 & 10 & 0 & member WBL 392  \\ %\nl

VCC 0350 & ImIV-V: & 305 & 0.6 & 5 & 0  \\ %\nl

VCC 0367 & ImV? & 2362 & 0.3 & 5 & 0 & p.w. N4270@1' \& $\Delta$v=5 km sec$^{-1}$  \\ %\nl

VCC 0381 & ImV & 481 & 0.8 & 4 & 0  \\ %\nl  BCD

VCC 0477 & ImV & 1866 & 0.9 & 4 & 0   \\ %\nl

VCC 0530 & ImIV-V & 1299 & 1.1 & 3 & 0 & p.w. IC 0783A@4'.9 \& $\Delta$v=92 km sec$^{-1}$\\ %\nl

VCC 0565 & ImIV & 877 & 0.5 &  7 & 0   \\

VCC 0584 & ImIV-V & 56 & 0.6 & 4 & 0 & p.w. VCC 0571@2'.9 \& $\Delta$v=991 km sec$^{-1}$ \\ %\nl

VCC 0628 & ImV: & --398 & 0.3  & 12 & 0  \\ %\nl

VCC 0826 & ImIV & 1505 & 1.3 & 3 & 0  \\ %\nl

VCC 0963 & Im: & 1866 &  0.4 & 2 & 0  \\ %\nl

VCC 1448 & ImIV or dE1 pec & 2583 & 1.7 & 3 & 0  \\ %\nl

VCC 1455 & ImIV & 1340 & 0.6 & 5 & 0  \\ %\nl

VCC 1468 & ImIV & 1233 & 0.9 & 4 & 0  \\ %\nl

VCC 1585 & ImIII-IV pec & 668 & 1.5 & 3 & 0 & IC 3522  \\ %\nl

VCC 1753 & ImIV & 737 & 0.6 & 5 & 0  \\ %\nl

VCC 1784 & ImV & 57 & 0.7 & 5 & 0  \\ %\nl

VCC 1816 & ImIV-V & 1006 & 1.0 & 7 & 0  \\

VCC 1822 & ImIV & 1012 & 0.6 & 10 & 0 &   \\ %\nl

VCC 1952 & ImIV & 1308 & 0.6 & 7 & 0  \\ %\nl

VCC 1992 & ImIV & 1010 & 1.1 & 4 & 0 & U7906  \\ %\nl

VCC 2034 & ImIV & 1500 & 0.8 & 10 & 1 & p.w. VCC 2037@3' \& $\Delta$v=459 km sec$^{-1}$ \\ %\nl

\end{tabular}
\end{minipage}
\end{table*}

%\hline

\begin{table*}
\begin{minipage}{150mm}
\caption{Possible HII companions of dwarf star-forming
galaxies: van Zee sample}
%\label{symbols}
\begin{tabular}{@{}ccccccl}
\hline
Name & Type & v$_{\odot}$ & a & A & N & Remarks \\ \hline
%LvZ
DDO210 & IB(s)m & --137 & 2.2 & 2 & 0 &   \\ %\nl

HARO43 & BCD? & 1912 & 0.5 & 12 & 0 &  \\ %\nl

*U10281 & Im: & 1080 & 1.8 &  5 & 0 & p.w. BH16110+1720@1'.5 \& $\Delta$v=5 km sec$^{-1}$ \\ %\nl

U1175 & Sm: & 728 & 1.1 & 10 & 0 &    \\ %\26nl

U11820 & Sm: & 1104 & 2.0 & 2 & 0 &   \\ %\nl

U191 & Sm & 1144 & 1.6 & 5  & 0 &   \\ %\nl

U2162 & IB(s)m & 1185 & 1.5 & 6 & 0 &    \\ %\nl

U2535 & Im: &  2968 & 0.9 & 21 & 0 &   \\ %\nl

U2684  & Im? & 350 & 1.8 & 4 & 0 &   \\ %\nl

U2984  & SBdm: & 1543 & 1.7 & 4 & 0 & p.w. NPM1G +13.0146@4' \& $\Delta$v=10 km sec$^{-1}$ \\ %\nl

U300  & Im: & 1346 & 1.3 & 5 & 0 &    \\ %\nl

U3050 & S? & 2147 & 0.9 & 15 & 0 &    \\ %\nl

U3174 & IAB(s)m: &  670 & 1.7 & 2 & 0 &  \\ %\nl

U3672  & Im & 994 & 1.2 & 5 & 0 &   \\ %\nl

U4660 & Sm: & 2203 & 1.2 & 4 & 0   \\ %\nl

U4762 & Im & 2026 & 1.0 & 12 & 0 &   \\ %\n26l

U521 & Im & 659 & 0.9 & 11 & 0 &  \\ %\nl

U5716 & Sm: & 1277 & 1.3 & 2 & 0 &   \\ %\nl

U5764 & IB(s)m: & 586 & 2.0 & 7 & 0 &   \\ %\nl

U5829 & Im & 629 & 4.7 & 1 & 0 & VV 794   \\ %\nl

U634 & SABm: & 2216 & 1.7 & 4 & 0 &   \\ %\nl

U7178 & IAB(rs)m: & 1339 & 1.4 & 4 & 0 &   \\ %\nl

U7300  & Im & 1210 & 1.4 & 6 & 0 &   \\ %\nl

U8024 &  IB(s)m IV-V & 376 & 3.0 & 4 & 0 &   \\ %\nl

U891 & SABm: & 643 & 2.3 & 2 & 0 &    \\ %\nl

U9128 & ImIV-V & 154 & 1.7 & 5 & 0 &   \\ %\nl

*U9762  & Sm: & 2273 & 1.0 & 4 & 0 & p.w. 2MASXi J1511566+323553@4'.2 \& $\Delta$v=43 km sec$^{-1}$ \\ %\nl

UA357 & Im & 1170 & 0.8 & 2 & 0 &   \\ %\nl

\end{tabular}
\end{minipage}
\end{table*}
%\hline
%SalzerJohn

\begin{table*}
\begin{minipage}{150mm}
\caption{Possible HII companions of dwarf star-forming
galaxies: Salzer sample}
%\label{symbols}
\begin{tabular}{@{}ccccccl}
\hline
Name & Type & v$_{\odot}$ & a & A & N & Remarks \\ \hline
IIZw40 & BCD; Sbc; merger HII &  789 & 0.6 & 2 & 0 & UGCA 116  \\ %\nl

IZw18 & BCD & 745 & 0.2 & 13 & 0  \\ %\nl

%DDO147 & 333 & 1.9 & none  \\ %\nl could not read H-alpha file into ds9

GR8 & ImV & 214 & 1.1 & 6 & 0 &   \\ %\nl

Leo A  & IBm: & 20 & 5.1 & 2 & 0   \\ %\nl

Mk324 & BCD & 1600 & 0.3 & 12 & 0 &   \\ %\nl

Mk328 & BCD/E & 1379 & 0.3 & 7 & 0 & UGCA 441  \\ %\nl

Mk36 & BCD & 646 & 0.3 & 7 & 0 & UGCA 225  \\ %\nl

Mk475 & BCD & 540 & 0.2 & 11 & 0 &   \\ %\nl

Mk5 & I? & 792 &  0.7 & 5 & 0 &   \\ %\nl

Mk600 & SBb; BCD & 1008 & 0.4 & 11 & 0 &  \\ %\nl

Mk750  & BCD & 754 & 0.4 & 6 & 0 &  \\ %\nl

Mk900 & BCD/E & 1146 & 0.8 & 9 & 0 &    \\ %\nl

UM133 & HII & 2098 & 1.0 & 5 & 0 &   \\ %\nl

UM323 & BCD? & 1799 & 0.5 & 7 & 0 &    \\ %\nl

UM408 & HII & 3598 & 0.2 & 23 & 0 &   \\ %\nl

UM40 & SB0? &  1344 & 0.6 & 5 & 0 &   \\ %\nl

*UM439 & compact & 1122 & 2.0 & 13 & 0 & UGC 6578; HI
companion?@1' \& $\Delta$v=27 km sec$^{-1}$
 \\ %\nl

UM461 & BCD/Irr; HII & 1007 & 0.4 & 11 & 0 &  \\ %\nl

UM462 & HII &  1051 & 0.6 & 7 & 0 & UGC 6850; two HI
companions@0'.1 \& $\Delta$v=144, 155 km sec$^{-1}$
 \\ %\nl

WAS 5 & BCD; HII & 1259 & 0.2 & 16 & 0 &   \\ %\nl
\hline

\end{tabular}
\end{minipage}
\end{table*}

The objects studied in this way are listed in Tables 1 to 4. The
tables give one of the names of the galaxy in the first column,
its morphological type in column 2, its heliocentric velocity and
semi-major axis (\emph{a}, in arcmin) in columns 3 and 4, the
distance from the galaxy searched for possible H$\alpha$
companions (A, in galactic radii, using the off-line or continuum
image) in column 5, and the number of possible H$\alpha$
companions identified in column 6. We add other names for the
galaxy in column 7 and sometimes we mention one or more possible
companions as pair with the galaxy (p.w. in the table) for which
we give the projected distance (see below). The redshift
v$_\odot$, the size of the semi-major axis $a$, and the
morphological type are the values listed in NED or in Hoffman
\etal (1987, 1989) or in Pustilnik \etal (1995). A few $a$ values
are major axis estimates from the images of the galaxies in DSS.
If the name itself is prefixed by an asterisk, we include below a
short discussion of this specific object. The different
sub-samples are separated in the table by horizontal lines.

Tables 1 to 4 indicate that only one DIG from our sample of 96
objects has an HII companion (VCC 2034, with VCC 2037). To this we
can add 12 galaxies with possible companions that do not show
H$\alpha$ emission found through the NED search. A few objects in
these Tables are worth a few additional comments:

\begin{itemize}
\item {\bf VCC 0340} This object is listed as the seventh galaxy
in the poor cluster WBL 392 in the catalog of White \etal (1999),
along with 11 other objects. The WBL catalog lists concentrations
of three or more galaxies brighter than m$_{pg}$=15.7 with a
surface overdensity of 21.54 (10$^{4/3}$). The redshift of the
proposed poor cluster is derived as the average of the redshifts
of the individual galaxies. Given its location and redshift, it is
clear that WBL 392 is a galaxy condensation within the Virgo
cluster, near its southern edge. Table 2 in White \etal lists a
few other similar condensations nearby.

\item {\bf VCC 0329} This object is listed as member in the same
poor cluster WBL 392 as VCC 0340. Note that while VCC 0340 is
classified as a BCD/merger, VCC 0329 is an LSB dG. The two dwarf
galaxies are only five arcmin ($\sim$24 kpc., projected) and 110
km sec$^{-1}$ apart, yet the first is copiously forming stars
while the second is not, although it has significant amounts of HI
(approximately 1/3 of the HI in VCC 0340 and with a similar line
width; Hoffman \etal 1987).

\item {\bf U10281} A search of NED in the neighborhood of this
galaxy reveals an apparent companion only 1'.5 away. This is
[BH98]16110+1720 listed with a redshift that differs only by 5 km
sec$^{-1}$ from that of U10281, originating from the study of the
Hercules supercluster kinematics by Barmby \& Huchra (1998).
However, the DSS image of the potential companion displayed by NED
shows nothing. It seems that the table listing the galaxies
studied by Barmby \& Huchra contained some typing mistakes that
were not corrected in the erratum to the paper, and were
propagated into NED.

\item {\bf U9762} This galaxy has two companions. One is listed in
Table 1 here, from a NED search. The two are discussed by Pisano
\& Wilcots (1999), where the companions were found through an HI
imaging survey with the VLA-D. The companion listed in Table 1 is
a compact dwarf galaxy with an extended HI envelope and with
H$\alpha$ emission. The other (northern) companion is an LSB
object with 7.1 10$^8$ M$_{\odot}$ of HI and a dynamical mass (cf.
Pisano \& Wilcots) higher than 10$^{10}$ M$_{\odot}$.

\item {\bf UM 439} NED carries a note about this object regarding
the presence of a possible HI companion. This is the result of
interpreting the disturbed and asymmetric HI profile measured by
Taylor \etal (1995). However, a later modification of this NED
note by C. Taylor, mentioned also in the original paper, indicates
that the disturbance could be the result of a tidal interaction,
even though the disturber was not readily identified. The contours
in their Figure 4 show that the HI distribution is centered on the
optical galaxy; the east-west distortion and the HI peak at 1110
km sec$^{-1}$ and 1.5 arcmin north-west of the optical galaxy
could represent the companion, though no optical counterpart is
readily discernable.

\end{itemize}

\section{Discussion}
We showed above that samples of dwarf, star-forming galaxies do
not show the presence of significant numbers of companions that
have contemporaneous star formation with the target galaxy.
Although the survey is not uniform and does not cover equal areas
around each galaxy, our composite look searches for neighbors in a
range of distances, both in the immediate neighborhood (up to five
galactic radii for half of the sample) and to within 20 radii or
more for $\sim$one quarter of the galaxies. The results for our
sample of 96 galaxies are indicative, but do not represent any
complete sample. To summarize, we found that one dwarf galaxy has
a possible H$\alpha$ companion and, through NED searches,
identified 12 other cases of possible companions that do not show
H$\alpha$ emission. In comparison, the majority of companions of
large galaxies in the Zaritsky \etal sample do show line emission.

The lack of detected H$\alpha$ companions shows that the
tidally-induced SF trigger is probably not operating in DIGs, or
is not important in such galaxies, some of which may be, as
already mentioned, ``first-time'' galaxies. This conclusion
confirms previous similar statements ({\it e.g.,} Mihos \etal
1997, Telles \& Melnick 1999, Telles \& Maddox 1999). We also
showed that the cluster tides are probably not relevant to the
question addressed here.

In fact, we can provide some counterexamples where one would
expect a strong tidal interaction from a massive companion, the
dwarf galaxy has considerable amounts of HI, yet no visible star
formation is detected. One such case is VCC 0367, an LSB dIrr (Im
V) that is only one arcmin and 5 km sec$^{-1}$ away from the large
S0 galaxy NGC 4270. VCC 0367 has considerable amounts of HI: 0.59
Jy km sec$^{-1}$ (Huchtmeier \& Richter 1989) and the 21-cm line
is fairly wide (w$_{50}$=98$\pm$10 km sec$^{-1}$; Bottinelli \etal
1990). The lack of any signs of interaction in this case, when one
would expect a strong tide to act between galaxies that are so
close to each other, indicates presumably a failure of the basic
assumption that a small angular distance coupled with a negligible
redshift difference implies necessarily that the two objects are
equidistant from us. Presumably one of the two galaxies is located
at a large distance from the other and has a strong peculiar
velocity that mimics the redshift of the other.

We are left in a somewhat difficult position in any attempt to
explain the pattern of SF we observe in DIGs. To recapitulate, we
showed that both high surface brightness (BCD) as well as LSB DIGs
exhibit SF. The star formation takes place mostly at the outer
boundaries of a DIG and not at its center (except for some BCDs).
It also tends to concentrate more to one side of a galaxy, rather
than being distributed with equal probability over the galaxy.
Unless we assume that near each such galaxy there is a dark matter
halo lacking HI and luminous stars, which provides the tidal force
to trigger the SF, we are compelled to abandon an attempt to
explain SF initiation by interactions.

The lack of external SF triggers implies that, in any explanation,
we must turn to internal phenomena to trigger the process, but as
we already mentioned in the introduction, large-scale internal
triggers, that could be relied upon to produce asymmetry (such as
density waves or a bar instability), are absent in DIGs. In this
context, we note that Hunter (1997) already concluded that SF is
largely a regional process and that in irregular galaxies it is
likely regulated by a combination of processes such as
gravitational instabilities, thermal pressure of gas in a disk,
and modification of the ISM by massive stars, as well as random
gas motions.

Lacking a plausible trigger for the SF process, we propose here a
new possibility, which is at this stage only a scenario, not a
full-power model. We call this the ``churning bag'' idea, which,
simply put, assumes that SF takes place when a mass of gas is
sloshing around within the dark matter (DM) halo of a dwarf
galaxy. The gas could find loci of standing waves probably at the
turn-around points, where the gas may be compressed and could
reach SF conditions. As this is a three-dimensional problem in a
possibly non-spherical (even triaxial) potential, many suitable
locations may exist in a DM halo. These may appear randomly
distributed over the face of a galaxy, but will tend to occur
preferentially near the edges of the gas mass, and there it would
be more likely to find locations that could form stars. This could
explain the finding of off-center star formation in DIGs (cf.
Brosch \etal 1998; Heller \etal 2000).

The question of gas oscillations in the gravitational potential of
a galaxy with subsequent star formation has been studied by Nelson
(1976) in the context of searching for a mechanism to explain the
``corrugations'' observed in the Galactic disk, and by de Zeeuw
(1984) in the context of the gas fate in a triaxial gravitational
potential. A related phenomenon, external shells in elliptical
galaxies, has been studied by Quinn (1984) and by Hernquist \&
Quinn (1987).

The problems of galactic-scale oscillations are similar, because
the observed peculiarities arise when the ``tracer'' material, be
it either gas or stars, adjusts itself to an external
gravitational potential. The difference is that gas is much more
dissipative than stars. In the case of shell galaxies, the
observed features are produced by the accretion of a low-mass
companion galaxy. The external shells are explained as being the
result of stars from the accreted object sloshing in the
gravitational potential of the elliptical galaxy. Each shell forms
at the turning points of the accreted stars. The simulations deal
mostly with the fate of stellar systems as the accreted
companions.

Weil \& Hernquist (1993) studied merger simulations of systems
consisting of gas and stars. The difference, with respect to
previous calculations, is the rapid separation of the gas from the
stars, with the gas settling in compact disks or rings near the
central region of the accretor galaxy at the bottom of the
gravitational potential. The stars form essentially a
collisionless fluid affected on longer timescales by dynamical
friction. One example Weil \& Hernquist presented in this context is
NGC 2685, the Spindle galaxy, an object showing traces of at least
two accretion/merger events. The more recent one resulted in a
luminous ring along the minor axis of the galaxy, in which SF
takes place at present (Eskridge \& Pogge 1997). H$\alpha$ images
of NGC 2685 show compact HII regions within the ring; these are
rather similar to the HII regions we observe in the DIGs we
studied and, as mentioned above, this star formation process does take
place in the outer regions of NGC 2685.

An oscillatory behavior of the star formation process, induced by
a mass exchange process between the disk and the halo of a galaxy,
has been studied by Korchagin \etal (1994). This, as noted also by
Hirashita \etal (2001), could explain the intermittent SF history
of many galaxies including our own. Note, though, that the
``oscillatory'' description relates in this case to the rate of SF
in a certain location, not to the spatial shifting of the SF
regions within a galaxy.

Sloshing, in the context of a galaxy-galaxy interaction, was
studied by Zeltwanger \etal (2000). They simulated the rapid
parabolic passage of a companion next to a target disk galaxy and
found that this could bring the center of mass (CoM) of the disk
to an offset position. The offset CoM would subsequently decay and
this would create a transient, one-armed spiral pattern. Given
that this is a 2D N-body simulation that does not include gas, we
would expect a more dramatic behavior from a gassy disk.

The sloshing of a stellar cluster within the halo of the dwarf
spheroidal galaxy in Ursa Minor was described by Kleyna \etal
(2003). They analyzed stellar radial velocities in this nearby
galaxy and found two kinematically-separated stellar populations.
This demonstrates that decoupled structures can survive for long
periods, a number of Gyrs, in the halo of a dwarf galaxy. In the
case of a dissipative mass of gas it is likely that the lifetime
of such an entity would be rather short, but presumably sufficient
to trigger a SF event near the edge of the galaxy at the
turn-around location.

We have established above that a sloshing mode of motion can take
place in a galaxy, and now propose a scenario for the sloshing
mode of star formation. We emphasize again that this description
is not based on a model calculation. Imagine a mass of gas
recently collected by a small dark-matter halo, similar to those
thought to exist around DIGs. The gas could originate from a
recent accretion event, or could be a concentrated blob ejected
from the galaxy itself and re-accreted by it. Before the gas fully
relaxes to a disk it is likely to oscillate within the
gravitational potential of the halo. These oscillations would
mostly tend to be radial and when a gas mass would reach its
apo-galactic point it will reside there longer than in regions
closer to the center of the potential. Moreover, the central
regions will have a $\sim$flat gravitational potential profile,
thus will not be conducive to much mass concentration. Adding to
this the lower cooling efficiency of metal-poor gas, which would
reduce the SFR to less than 1/3 of that in metal-rich gas (c.f.
Gerritsen \& de Blok 1999), implies that the outer regions of the
galaxy will tend to form stars more efficiently than other
locations in it, as indeed our studies have found. This star
formation process would then be similar to the SF trigger
discussed by Elmegreen (1998) as the ``collect and collapse''
mode, in which the gas accumulates in a dense ridge that collapses
gravitationally into dense cores.

Note that we did not discuss here how the gas became trapped in
the DM halo. This may be the result of slow cooling of gas in the
halo of the proto-galaxy (cooling flow), or gas could be accreted
from intergalactic space, or it could even be torn off another
proto-galaxy. As already mentioned above, this could even be gas
ejected from the galaxy and now on an infalling trajectory. The
presence of gas is an \emph{a-priori} requirement for recent star
formation, and there might be conditions where it could begin to
form stars without external influences; this is the main point of
the present paper.

\section{Conclusions}
We analyzed published samples of galaxies with companions and
showed that there does not seem to be a strong link between the
distance from a companion to the parent galaxy and the strength of
the star burst detected in a galaxy. We also demonstrated that in
a composite sample of dwarf star-forming galaxies there are few
star-forming companion galaxies. These findings support the view
that galaxy-galaxy tidal interactions, particularly the long-range
encounters proposed by Icke (1985) as one of the possible star
formation triggers, are not very much involved in activating the
SF process in dwarf galaxies. It is possible that some internal
triggers, such as the sloshing of gas in a dark-matter halo
followed by apo-galactic gas accumulation at the turn-around
locations, provide such primary triggers of SF. This could be the
case for some of the star-forming and apparently isolated
galaxies, such as those in our sample. In other objects, the SF
could be secondary, triggered by SNe, stellar winds, etc. Whether
this is indeed the case could be revealed by detailed studies of
the star formation history of these objects.

\section*{Acknowledgements}
NB is grateful to Ed Salpeter, Lyle Hoffman, and Simon Pustilnik
for discussions on the topic of this paper and acknowledges Liese
van Zee and John Salzer who provided images of their galaxies.
Astronomical research at the Wise Observatory is supported by a
grant from the Israel Science Foundation. NB acknowledges support
from the Austrian Friends of Tel Aviv University. Mr. Ilia
Anshelevitz helped with the H$\alpha$ image reduction of the full
Virgo sample of BCD galaxies. We gladly acknowledge the intensive
use of NED for the performance of this study and the constructive
remarks of an anonymous referee.

\newpage

% ----------------------------------------------------------------
%\bibliographystyle{amsplain}
%\bibliography{}
\section*{References}
%\begin{description}

\end{document}